\documentclass[
reprint,
showpacs,
showkeys,
 amsmath,amssymb,
 aps,
prd,
floatfix,
]{revtex4-1}
\usepackage{color}
\usepackage{slashed}
\usepackage{graphicx}
\usepackage{dcolumn}
\usepackage{bm}

\begin{document}


\title{QCD phase diagram from chiral symmetry restoration: analytic approach at high and low temperature using the Linear Sigma Model with Quarks}

\author{Alejandro Ayala$^{1,2}$, Sa\'ul Hern\'andez-Ortiz$^1$ and L. A. Hern\'andez$^{1,2}$}
\affiliation{$^1$Instituto de Ciencias
  Nucleares, Universidad Nacional Aut\'onoma de M\'exico, Apartado
  Postal 70-543, M\'exico Distrito Federal 04510,
  M\'exico.\\
  $^2$Centre for Theoretical and Mathematical Physics, and Department of Physics,
  University of Cape Town, Rondebosch 7700, South Africa.}
  

\begin{abstract}

We use the linear sigma model with quarks to study the QCD phase diagram from the point of view of chiral symmetry restoration. We compute the leading order effective potential for high and low temperatures and finite quark chemical potential, up to the contribution of the ring diagrams to account for the plasma screening effects. We fix the values of the model couplings using physical values for the input parameters such as  the vacuum pion and sigma masses, the critical temperature at vanishing quark chemical potential and the conjectured end point value of the baryon chemical potential of the transition line at vanishing temperature. We also make the analysis for the same input parameters but with vanishing pion mass. We find that the critical end point (CEP) is located at low temperatures and high quark chemical potentials $(315<\mu^{\text{CEP}}<349\ {\mbox{MeV}},18< T^{\text{CEP}}<45\ {\mbox{MeV}})$.
\end{abstract}

\maketitle


\section{\label{sec:level1}Introduction}

Among the important subjects of study in the realm of high-energy/nuclear physics, both from the theoretical and experimental points of view, are the properties of strongly interacting matter under extreme conditions of temperature and baryon density. Of particular interest is the location of the Critical End Point (CEP) in the QCD phase diagram. To this aim, the STAR BES-I program has recently analyzed collisions of heavy-nuclei in the energy range 200 GeV $ > \sqrt{s_{NN}} > $ 7.7 GeV~\cite{BESI}. Future experiments~\cite{BESII,FAIR,NICA} will keep on conducting a thorough exploration of the transition from confined/chiral-symmetry broken hadron matter to the deconfined/chiral-symmetry restored state, varying the temperature and baryon density by changing the collision energy down to about $\sqrt{s_{NN}}\simeq 5$ GeV and the system size in hadron and heavy-ion reactions. From the theoretical side, efforts to locate the CEP employing a variety of techniques such as Schwinger-Dyson equations, finite energy sum rules, functional renormalization methods, holography, and effective models, have produced a wealth of results~\cite{values,Ayala-Dominguez,Xin,Fischer,Lu,Shi,Contrera,Cui,Datta,Knaute,Antoniou,Rougemont} ranging from low to large values of the baryon chemical potential ($\mu_B$) and temperature ($T$). Recent lattice QCD (LQCD) analyses~\cite{lattice} have resorted to using the imaginary baryon chemical potential technique, to later extrapolate to real values, to study the chiral transition near the $T$-axis. Albeit with still large uncertainties, this technique has shown that the transition keeps being a smooth crossover~\cite{Bellwiede}. The Taylor expansion LQCD technique has also been employed to restrict the CEP's location to values $\mu_B/T>2$ for the temperature range 135 MeV $< T <$ 155 MeV. Its location for temperatures larger than $0.9\ T^c(\mu_B=0)$ seems to also be highly disfavored~\cite{Bazavov} (see also~\cite{Schmidt}).

Effective models have proven to be useful tools to gain insight into the phase structure of QCD. Given the dual nature of the QCD phase transition, at least for low values of $\mu_B$, one can ask whether models that incorporate both chiral symmetry breaking and deconfinement are better suited to describe the transition features. However, since LQCD results show that for 2+1 light flavors, the crossover chiral and deconfinenent transitions are indistinguishable~\cite{latticeTc}, one may resort to a simplified analysis whereby one or the other feature is emphasized. Recently, we have made use of the linear sigma model coupled to quarks~\cite{Ayala1,Ayala2}. We have shown that this tool can be successfully employed provided one accounts for the screening properties of the plasma, which makes the analysis effectively go beyond the mean field approximation, and one finds the values of the couplings from the physical values of the model parameters. 

In this work we use the linear sigma model coupled to quarks, including the plasma screening effects, to explore the effective QCD phase diagram from the point of view of chiral symmetry restoration. Our strategy is to fix the coupling constants using the physical values of the model parameters, such as the vacuum pion and sigma masses, the critical temperature $T^c$ at $\mu_B=0$ and the conjectured end point value of $\mu_B\ (\simeq 1$ GeV) of the transition line at $T=0$. For the present purposes we compute an analytical, leading order in $T$ approximation for the effective potential, both at high and low temperatures, for finite values of the baryon chemical potential. We show that this strategy can be used to locate the CEP. The work is organized as follows: In Sec.~\ref{sec:level2}, we introduce the linear sigma model coupled to quarks. In Sec.~\ref{sec:level3} we compute the effective potential up to the contribution of the ring diagrams. We work out the high and low temperature analytical approximation for the effective potential and show explicitly how in the high temperature domain, the ring diagrams contribution cures the non-analyticities that appear at one-loop order. In Sec.~\ref{sec:level7} we spell out the conditions that give rise to the equations to find the values of the model coupling constants. In Sec.~\ref{sec:level8} we use these couplings to compute the critical $T$ and $\mu_B$ values that define the transition curves and locate the CEP. We finally summarize and conclude in Sec~\ref{sec:level9}. We reserve for the appendices the calculation details for the boson and fermion contributions to the one-loop effective potential. In a sequel, to be reported elsewhere, we will study the case where the analytical approximation is extended to cover a larger set of possible $\mu_B$ and $T$ values as well as to include the case where the couplings are allowed to bear the dependence on $\mu_B$ and $T$.

\section{\label{sec:level2}Linear Sigma Model coupled to quarks}

In order to explore the QCD phase diagram, we study the restoration of chiral symmetry using an effective model that accounts for the physics of spontaneous symmetry breaking at finite temperature and density, the \textit{Linear Sigma Model}. In order to account for the fermion degrees of freedom around the phase transition, we also include quarks in this model. The Lagrangian for the linear sigma model when the two lightest quarks are included is given by
\begin{eqnarray}
   \mathcal{L}&=&\frac{1}{2}(\partial_\mu \sigma)^2  + \frac{1}{2}(\partial_\mu \vec{\pi})^2 + \frac{a^2}{2} (\sigma^2 + \vec{\pi}^2) - \frac{\lambda}{4} (\sigma^2 + \vec{\pi}^2)^2 \nonumber \\
   &+& i \bar{\psi} \gamma^\mu \partial_\mu \psi -g\bar{\psi} (\sigma + i \gamma_5 \vec{\tau} \cdot \vec{\pi} )\psi ,
\label{lagrangian}
\end{eqnarray}
where $\psi$ is an SU(2) isospin doublet, $\vec{\pi}=(\pi_1, \pi_2, \pi_3 )$ is an isospin triplet and $\sigma$ is an isospin singlet. $\lambda$ is the boson's self-coupling and $g$ is the fermion-boson coupling. $a^2>0$ is the mass parameter. 

To allow for an spontaneous breaking of symmetry, we let the $\sigma$ field to develop a vacuum expectation value $v$
\begin{equation}
   \sigma \rightarrow \sigma + v,
\label{shift}
\end{equation}
which can later be taken as the order parameter of the theory.  After this shift, the Lagrangian can be rewritten as
\begin{eqnarray}
   {\mathcal{L}} &=& -\frac{1}{2}[\sigma \partial_{\mu}^{2}\sigma]-\frac{1}
   {2}\left(3\lambda v^{2}-a^{2} \right)\sigma^{2}\nonumber \\
   &-&\frac{1}{2}[\vec{\pi}\partial_{\mu}^{2}\vec{\pi}]-\frac{1}{2}\left(\lambda v^{2}- a^2 \right)\vec{\pi}^{2}+\frac{a^{2}}{2}v^{2}\nonumber \\
  &-&\frac{\lambda}{4}v^{4} + i \bar{\psi} \gamma^\mu \partial_\mu \psi 
  -gv \bar{\psi}\psi + {\mathcal{L}}_{I}^b + {\mathcal{L}}_{I}^f,
  \label{lagranreal}
\end{eqnarray}
where ${\mathcal{L}}_{I}^b$ and  ${\mathcal{L}}_{I}^f$ are given by
\begin{eqnarray}
  {\mathcal{L}}_{I}^b&=&-\frac{\lambda}{4}\Big[(\sigma^2 + (\pi^0)^2)^2\nonumber \\ 
  &+& 4\pi^+\pi^-(\sigma^2 + (\pi^0)^2 + \pi^+\pi^-)\Big],\nonumber \\
  {\mathcal{L}}_{I}^f&=&-g\bar{\psi} (\sigma + i \gamma_5 \vec{\tau} \cdot \vec{\pi} )\psi.
  \label{lagranint}
\end{eqnarray}

Equation~(\ref{lagranint}) describes the interactions among the $\sigma$, $\vec{\pi}$ and $\psi$ fields after symmetry breaking. From Eq.~(\ref{lagranreal}) one can see that the sigma, the three pions and the quarks have masses given by
\begin{eqnarray}
  m^{2}_{\sigma}&=&3  \lambda v^{2}-a^{2},\nonumber \\
  m^{2}_{\pi}&=&\lambda v^{2}-a^{2}, \nonumber \\
  m_{f}&=& gv,
\label{masses}
\end{eqnarray}
respectively.

In order to determine the chiral symmetry restoration conditions as function of temperature and quark chemical potential, we study the behavior of the effective potential, which we now proceed to deduce in detail.

\section{\label{sec:level3}Effective potential}

Chiral symmetry restoration can be identified by means of the finite temperature and density effective potential, which in turn is computed order by order. In this work we include the classical potential or tree-level contribution, the one-loop correction both for boson and fermions and the ring diagrams contribution, which accounts for the plasma screening effects. 

The tree level potential is given by

\begin{equation}
    V^{\text{tree}}(v)=-\frac{a^2}{2}v^2+\frac{\lambda}{4}v^4,
    \label{treelevel}
\end{equation}
whose minimum is given by

\begin{equation}
    v_0=\sqrt{\frac{a^2}{\lambda}},
\end{equation}
since $v_0\neq 0$, we notice that the symmetry is spontaneously broken. We also notice that

\begin{equation}
    \frac{d^2 V^{\text{tree}}}{dv^2}=3\lambda v^2-a^2=m_\sigma^2,
    \label{curvature}
\end{equation}
which means that the curvature of the classical potential is equal to the sigma mass squared. This property is maintained even when corrections due to finite temperature and density are included in the effective potential.

However, in order to make sure that the quantum corrections at finite temperature and density maintain the general properties of the effective potential, we need to add counter-terms $\delta a^2$ and $\delta \lambda$ to the bare constants $a^2$ and $\lambda$, respectively, and write 

\begin{align}
	V^{\text{tree}}&=-\frac{a^2}{2}v^2+\frac{\lambda}{4}v^4 \nonumber \\
    &\rightarrow -\frac{(a^2+\delta a^2)}{2}v^2+\frac{(\lambda+\delta \lambda)}{4}v^4.
    \label{newtree}
\end{align}
These counter-terms are needed to make sure that the phase transition at the critical temperature $T_c$ for $\mu_B=0$ is second order and that this transition is first order at the critical baryon density $\mu_B^c=0$ for $T=0$. We will come back to these conditions when we introduce the analysis to determine the parameters of the model.

To include quantum corrections at finite temperature and density, we work within the imaginary-time formalism of thermal field theory. The general expression for the one-loop boson contribution can be written as

\begin{equation}
    V^{(1)\text{b}}(v,T)=T\sum_n\int\frac{d^3k}{(2\pi)^3} \ln D(\omega_n,\vec{k})^{1/2},
    \label{1loopboson}
\end{equation}
where

\begin{equation}
D(\omega_n,\vec{k})=\frac{1}{\omega_n^2+k^2+m_b^2},
\end{equation} 
is the free boson propagator with $m_b^2$ being the square of the boson's mass and $\omega_n=2n\pi T$ the Matsubara frequencies for boson fields. 

For a fermion field with mass $m_f$, the general expression for the one-loop correction at finite temperature and quark chemical potential $\mu_q$ is

\begin{equation}
    V^{(1)\text{f}}(v,T,\mu_q)=-T\sum_n\int\frac{d^3k}{(2\pi)^3} \text{Tr}[\ln S(\tilde{\omega}_n-i\mu_q,\vec{k})^{-1}],
    \label{1loopfermion}
\end{equation}
where

\begin{equation}
S(\tilde{\omega}_n,\vec{k})=\frac{1}{\gamma_0 \tilde{\omega}_n+\slashed{k}+m_f},
\end{equation}
is the free fermion propagator and $\tilde{\omega}_n=(2n+1)\pi T$ are the Matsubara frequencies for fermion fields. 

The ring diagrams term is given by

\begin{eqnarray}
    V^{\text{Ring}}(v,T,\mu_q)&=&\frac{T}{2}\sum_n\int\frac{d^3k}{(2\pi)^3}\nonumber\\
    &\times&\ln (1+\Pi(m_b,T,\mu_q)D(\omega_n,\vec{k})),
    \label{rings}
\end{eqnarray}
where $\Pi(m_b,T,\mu_q)$ is the boson's self-energy.

\begin{figure}[t]
\begin{center}
\includegraphics[scale=0.64]{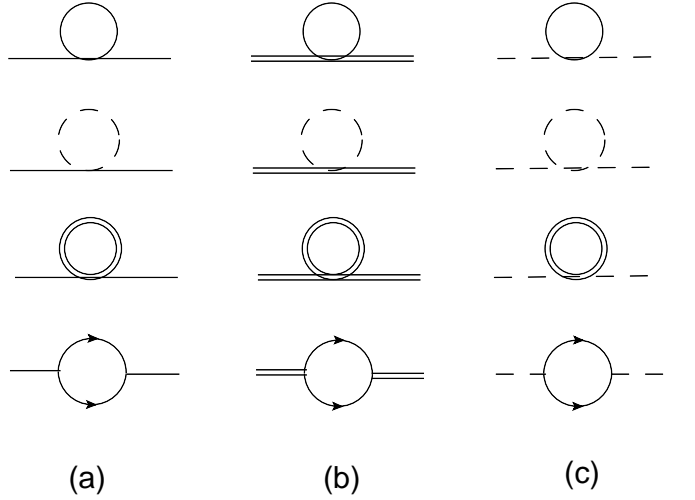}
\end{center}
\caption{Feynman diagrams contributing to the one loop bosons' self-energies. The dashed line denotes the charged pion, the continuous line is the sigma, the double line represents the neutral pion and the continuous line with arrows represents the fermions.}
\label{fig1}
\end{figure}

\subsection{\label{sec:level4}Self-energy.}

We start by computing the self-energy for one boson field. For this purpose we need to include all the contribution from the Feynman rules in Eq.~(\ref{lagranint}). The diagrams representing the bosons' self-energy are depicted in Fig.~\ref{fig1}. Each boson has a self-energy with two kinds of terms, one corresponds to a loop made by a boson field and other one corresponding to a loop made by a fermion anti-fermion pair. Therefore, the self-energy is written as

\begin{equation}
    \Pi(T,\mu_q)=\sum_{i=\sigma, \pi^0,\pi^\pm}\Pi_{\text{i}}(T)+\sum_{j=u,d}\Pi_{\text{j}}(T,\mu_q),
\end{equation}
where
\begin{align}
    \Pi_\sigma(T)&=\frac{\lambda}{4}[12 I(\sqrt{m_\sigma^2+\Pi_\sigma})+4I(\sqrt{m^2_{\pi^0}+\Pi_{\pi^0}}) \nonumber \\
    &+8I(\sqrt{m^2_{\pi^\pm}+\Pi_{\pi^\pm}})],\nonumber \\
    \Pi_{\pi^\pm}(T)&=\frac{\lambda}{4}[4 I(\sqrt{m_\sigma^2+\Pi_\sigma})+4I(\sqrt{m^2_{\pi^0}+\Pi_{\pi^0}}) \nonumber \\
    &+16I(\sqrt{m^2_{\pi^\pm}+\Pi_{\pi^\pm}})],\nonumber \\
    \Pi_{\pi^0}(T)&=\frac{\lambda}{4}[4 I(\sqrt{m_\sigma^2+\Pi_\sigma})+12I(\sqrt{m^2_{\pi^0}+\Pi_{\pi^0}}) \nonumber \\
    &+8I(\sqrt{m^2_{\pi^\pm}+\Pi_{\pi^\pm}})],
\end{align}
with
\begin{equation}
    I(x)=\frac{1}{2\pi^2}\int dk \frac{k^2}{\sqrt{k^2+x}} n(\sqrt{k^2+x}),
\end{equation}
and $n(x)$ being the Bose-Einstein distribution.


The leading temperature approximation to the boson self-energy is given by 

\begin{align}
    \Pi_\sigma(T)&=\Pi_{\pi^\pm}(T)=\Pi_{\pi^0}(T)=\frac{\lambda}{4}[24I(0)]\nonumber \\
    &=\frac{6\lambda}{2\pi^2}\int dk k \frac{1}{e^{k/T}-1}=\frac{\lambda T^2}{6}.
    \label{selfenergyboson}
\end{align}

This approximation, where the boson's mass is neglected with respect to the temperature, is a good approximation around the phase transition where the boson's mass (including its thermal correction) vanishes, namely, $m_i^2+\Pi_i=0$. 

On the other hand, the fermion contribution is given by

\begin{align}
    \Pi_j(T,\mu_q)&=-g^2 T\sum_n\int \frac{d^3k}{(2\pi)^3}\text{Tr}[S(\tilde{\omega}_n-i\mu_q,\vec{k},m_f)\nonumber \\
    &\times S(\tilde{\omega}_n-i\mu_q-\tilde{\omega}_m,\vec{k}-\vec{p},m_f)].
    \label{selfEF}
\end{align}

Equation~(\ref{selfEF}) can be computed without resorting to assuming a hierarchy between $T$ and $\mu_q$. Also, since we work close to the phase transition, we take $m_f=0$. The fermion self-energy contribution becomes

\begin{equation}
 \Pi_j(T,\mu_q)=-N_cg^2\frac{T^2}{\pi^2}[\text{Li}_2(-e^{\mu_q/T})+\text{Li}_2(-e^{-\mu_q/T})].
 \label{selfenergyfermion}
\end{equation}

With Eqs.~(\ref{selfenergyboson}) and~(\ref{selfenergyfermion}), the total self-energy for one boson is

\begin{align}
    \Pi(T,\mu_q)=&-N_fN_cg^2\frac{T^2}{\pi^2}[\text{Li}_2(-e^{\mu_q/T})+\text{Li}_2(-e^{-\mu_q/T})] \nonumber \\
    &+\frac{\lambda T^2}{2}.
    \label{fullselfenergy}
\end{align}

With the boson self-energy at hand we can study the properties of the effective potential. In order to work with analytical expressions we turn to study two cases: first the high temperature approximation, {\it i.e.} $T\gg m_b, \ \mu_q$ and then the low temperature approximation {\it i.e.} $T\ll m_b, \ \mu_q$. In the following we compute explicitly both regimes.

\subsection{\label{sec:level5}High temperature approximation}

For small $\mu_B$ and the transition temperature for chiral symmetry restoration found by LQCD computations~\cite{latticeTc}, we observe that $T$ is the largest of the energy scales. Therefore, a high temperature approximation is suited to study the chiral symmetry restoration. Let's start from Eq.~(\ref{1loopboson}), the one-loop correction for boson fields. The first step is to compute the sum over Matsubara frequencies. On doing so we obtain

\begin{align}
    V^{(1)\text{b}}(v,T)&=\frac{1}{2\pi^2}\int dk \ k^2\Big \{ \frac{\sqrt{k^2+m_b^2}}{2}\nonumber \\
    &+T\ln \Big(1-e^{-\sqrt{k^2+m_b^2}/T} \Big) \Big \}.
    \label{V1bwosum}
\end{align}

Notice that Eq.~(\ref{V1bwosum}) has two pieces, the first one is the \textit{vacuum} contribution and the second one is the \textit{matter} contribution, namely, the $T$-dependent correction. In order to compute the vacuum term, we need to regularize and renormalize the former. For this purpose, we employ dimensional regularization and the Minimal Subtraction scheme (MS), with the renormalization scale $\tilde{\mu}=e^{-1/2}a$. For the matter term, we take the approximation $m_b/T\ll 1$ and we include only the most dominant terms (for more details see Appendix~A). Taking all this into account, the one-loop contribution to the effective potential from boson fields is given by

\begin{align}
    V_{\text{HT}}^{(1)\text{b}}(v,T)=&-\frac{m_b^4}{64\pi^2}\Big[ \ln \Big( \frac{4\pi a^2}{m_b^2}\Big)-\gamma_E+\frac{1}{2} \Big]\nonumber \\
    &-\frac{m_b^4}{64\pi^2}\ln \Big( \frac{m_b^2}{(4\pi T)^2}\Big)-\frac{\pi^2 T^4}{90}\nonumber \\
    &+\frac{m_b^2 T^2}{24}-\frac{m_b^3 T}{12\pi}.
    \label{final1loopb}
\end{align}

For the case of the fermion one-loop contribution, we follow the procedure outlined for the boson case. Thus, we start by computing the sum over the Matsubara frequencies to obtain

\begin{align}
    V^{(1)\text{f}}(v,T,\mu_q)&=-\frac{1}{\pi^2}\int dk \ k^2\Big \{ \sqrt{k^2+m_f^2}\nonumber \\
    &-T\ln \Big(1-e^{-(\sqrt{k^2+m_b^2}-\mu_q)/T} \Big) \nonumber \\
    &-T\ln \Big(1-e^{-(\sqrt{k^2+m_b^2}+\mu_q)/T} \Big)\Big \}.
    \label{V1fwosum}    
\end{align}

As for the boson case, we find that Eq.~(\ref{V1fwosum}) contains two pieces, one corresponding to the vacuum contribution and the other one to the matter contribution. The latter has the contribution of the quark chemical potential and for this reason we now have two terms corresponding to the particle and the anti-particle contributions. The vacuum contribution is computed exactly in the same manner for the boson case. For the matter term, we compute the integral in momentum taking into account the approximation where $m_f/T\ll 1$ and $\mu_q/T<1$, and we consider only the dominant terms (for more details see Appendix~B). After we compute the momentum integral in Eq.~(\ref{V1fwosum}) we get

\begin{align}
    V_{\text{HT}}^{(1)\text{f}}(v,T)&=\frac{m_f^4}{16\pi^2}\Big[ \ln \Big( \frac{4\pi a^2}{m_f^2}\Big)-\gamma_E+\frac{1}{2} \Big]\nonumber \\
    &+\frac{m_f^4}{16\pi^2}\Big[\ln \Big( \frac{m_f^2}{(4\pi T)^2}\Big)
    -\psi^0\Big( \frac{1}{2}+\frac{\text{i}\mu}{2\pi T} \Big)\nonumber \\
    &-\psi^0\Big( \frac{1}{2}-\frac{\text{i}\mu}{2\pi T} \Big)\Big]-8m_f^2T^2\Big[ \text{Li}_2(-e^{\mu_q/T})\nonumber \\
    &+\text{Li}_2(-e^{-\mu_q/T}) \Big]+32T^4\Big[ \text{Li}_4(-e^{\mu_q/T})\nonumber \\
    &+\text{Li}_4(-e^{-\mu_q/T}) \Big].
    \label{final1loopf}
\end{align}

In order to go beyond the mean field (one-loop) approximation, we need to consider the plasma screening effects. These can be accounted for by means of the ring diagrams. Since we are working in the high temperature approximation, we notice that the lowest Matsubara mode is the most dominant term~\cite{LeBellac}. Therefore we do not need to compute the other modes and Eq.~(\ref{rings}) becomes

\begin{align}
    V^{\text{Ring}}(v,T,\mu_q)&=\frac{T}{2}\int\frac{d^3k}{(2\pi)^3}\ln (1+\Pi(T,\mu_q)D(\vec{k}))\nonumber \\
    &=\frac{T}{4\pi^2}\int dk \ k^2 \Big \{ \ln(k^2+m_b^2+\Pi(T,\mu_q))\nonumber \\
    &-\ln(k^2+m_b^2) \Big\}.
    \label{rings2}    
\end{align}
From Eq.~(\ref{rings2}), we see that both integrands are almost the same except that one is modified by the self-energy and the other one is not. Thus, after integration, we obtain that the ring diagrams contribution is

\begin{equation}
    V^{\text{Ring}}(v,T,\mu_q)=\frac{T}{12\pi}(m_b^3-(m_b^2+\Pi(T,\mu_q))^{3/2}).
    \label{finalrings}
\end{equation}

With these pieces at hand, we can write the effective potential up to the ring diagrams contribution in the high temperature approximation. This is given by

\begin{align}
    V_{\text{HT}}^{\text{eff}}(v,T,\mu_q)&=-\frac{(a^2+\delta a^2)}{2}v^2
    +\frac{(\lambda+\delta \lambda)}{4}v^4\nonumber \\
    &+\sum_{b=\sigma,\bar{\pi}}\Big\{-\frac{m_b^4}{64\pi^2}\Big[ \ln \Big( \frac{ a^2}{4\pi T^2}\Big)-\gamma_E+\frac{1}{2} \Big]\nonumber \\
    &-\frac{\pi^2 T^4}{90}+\frac{m_b^2 T^2}{24}-\frac{(m_b^2+\Pi(T,\mu_q))^{3/2} T}{12\pi}\Big\}\nonumber \\
    &+\sum_{f=u,d}\Big\{\frac{m_f^4}{16\pi^2}\Big[ \ln \Big( \frac{ a^2}{4\pi T^2}\Big)-\gamma_E+\frac{1}{2}\nonumber \\
    &-\psi^0\Big( \frac{1}{2}+\frac{\text{i}\mu_q}{2\pi T} \Big)-\psi^0\Big( \frac{1}{2}-\frac{\text{i}\mu_q}{2\pi T} \Big)\Big]\nonumber \\
    &-8m_f^2T^2\Big[ \text{Li}_2(-e^{\mu_q/T})+\text{Li}_2(-e^{-\mu_q/T}) \Big]\nonumber \\
    &+32T^4\Big[ \text{Li}_4(-e^{\mu_q/T})+\text{Li}_4(-e^{-\mu_q/T}) \Big]\Big\}.
    \label{finalHTpotential}
\end{align}

Notice that the potentially dangerous pieces coming from linear or cubic powers of the boson mass, that could become imaginary for certain values of $v$, are removed or replaced by the contribution of the ring diagrams~\cite{DJ}.

\subsection{\label{sec:level6}Low temperature approximation}

To have access to the region in the QCD phase diagram where $\mu_B$ is large and $T$ is small, we compute the effective potential in the approximation where $T$ is the soft scale in the system. We call this the low temperature approximation. The approximation is applied both to the contribution of boson and fermion fields. 

In the case of boson fields, we include a boson chemical potential. We relate this to the energy required to add or remove one boson to the system. We associate this term to the description of high density in the analysis, in other words, the bosons' chemical potential $\mu_b$, is related to the conservation of an average number of particles and not to a conserved charge. The introduction of the boson's chemical potential is used to account for the possible onset of meson condensates as the quark chemical potential increases. This phenomenon has been described since long ago in the context of processes taking place in the core of neutron stars, where an excess of negative pions appears when the electron chemical potential approaches the pion rest mass~\cite{Bahcall}. In the present context, since the relevant interactions are between mesons and quarks, an excess of pions is bound to appear when the quark chemical potential approaches the pion mass. Therefore, the one-loop contribution for boson fields after the sum over Matsubara frequencies is

\begin{align}
    V_{\text{LT}}^{(1)\text{b}}(v,T,\mu_b)&=\frac{1}{2\pi^2}\int dk \ k^2\Big \{ \sqrt{k^2+m_b^2}\nonumber \\
    &+2T\ln \Big(1-e^{-(\sqrt{k^2+m_b^2}-\mu_b)/T} \Big) \Big \}.
    \label{V1bwosumLT}
\end{align}
In this approximation, it is not necessary to compute the vacuum and matter contributions separately, in fact the full expression can be computed at once. In this work, we follow the procedure used in Ref.~\cite{chilenos}. The general idea consists on developing a Taylor series around $T=0$ of the following expression

\begin{equation}
    V_{\text{LT}}^{(1)\text{b}}(v,T,\mu_b)=\int_{\frac{\mu_b-m_b}{T}}^\infty V_0^{\text{b}}(v,\mu_b+xT)h_B(x)dx,
    \label{1lLT}
\end{equation}
where $h_B(x)$ is the first derivative of the Bose-Einstein distribution and $V_0^{\text{b}}(v,\mu_b+xT)$ is the one-loop boson contribution evaluated at $T=0$, which is given explicitly by 
\begin{align}
    V_{0}^{(1)\text{b}}(v,\mu_b)&=-\frac{m_b^4}{64\pi^2}\Big[ \ln\Big( \frac{4\pi a^2}{(\mu_b+\sqrt{\mu_b^2-m_b^2})^2} \Big)\nonumber \\
    &-\gamma_E+\frac{1}{2}\Big]+\frac{\mu_b\sqrt{\mu_b^2-m_b^2}}{96\pi^2}(2\mu_b^2-5m_b^2).
    \label{V1bLT}
\end{align}

Notice that the one loop contribution from boson fields in the limit $T=0$, that appears in Eq.~(\ref{1lLT}), is evaluated at $\mu_b \rightarrow \mu_b+xT$. Then the expression of one-loop matter contribution from one boson field in the low temperature approximation becomes

\begin{align}
    V_{\text{LT}}^{(1)\text{b}}(v,T,\mu_b)=&V_0^{\text{b}}(v,\mu_b+xT)\Big |_{T=0}\nonumber \\
    &+\frac{\pi^2 T^2}{12}\frac{\partial^2}{\partial (xT)^2}V_0^{\text{b}}(v,\mu_b+xT)\Big |_{T=0} \nonumber \\
    &+\frac{7\pi^4 T^4}{1260}\frac{\partial^4}{\partial (xT)^4}V_0^{\text{b}}(v,\mu_b+xT)\Big |_{T=0}.
    \label{1loopBLT}
\end{align}
For more details see Appendix~C.

For fermion fields, we start from Eq.~(\ref{V1fwosum}), such that we implement the low temperature approximation in the same way as we did for boson fields. We now develop a Taylor series around $T=0$ of the following expression

\begin{figure}[t]
\begin{center}
\includegraphics[scale=0.48]{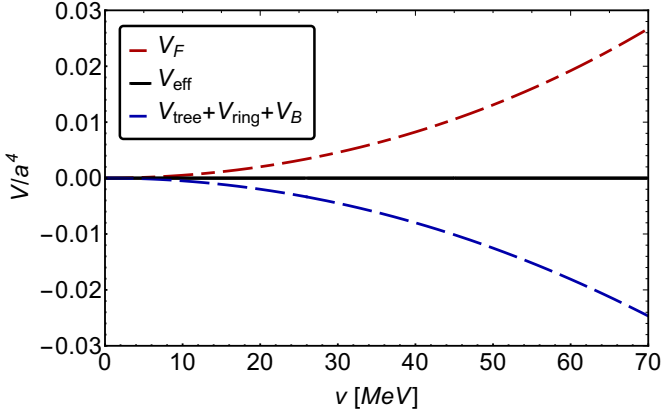}
\end{center}
\caption{Fermion and boson contributions to the effective potential at the phase transition at high temperature near the minimum at $v=0$. Notice that the sum of the two contributions offset each other making the potential to be flat. This is tantamount of a second order phase transition.}
\label{fig2}
\end{figure}

\begin{figure}[bh!]
\begin{center}
\includegraphics[scale=0.48]{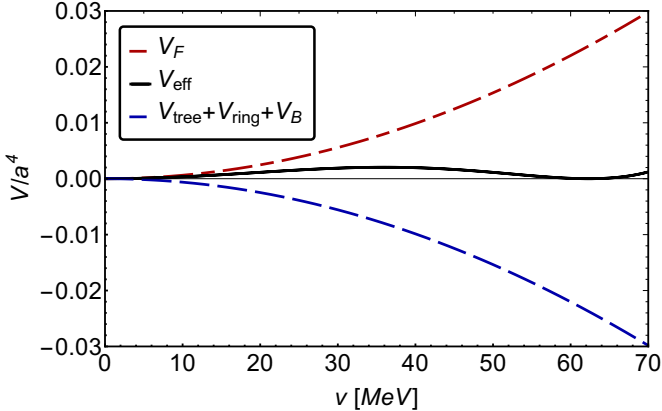}
\end{center}
\caption{Fermion and boson contributions to the effective potential at the phase transition for low temperature. Notice that the sum of the two contributions produce a barrier between each of the two degenerate minima at the transition. This is tantamount of a first order phase transition.}
\label{fig3}
\end{figure}

\begin{equation}
    V_{\text{LT}}^{(1)\text{f}}(v,T,\mu_q)=\int_{\frac{\mu_q-m_f}{T}}^\infty V_0^{\text{f}}(v,\mu_q+xT)h_F(x)dx,
    \label{1lLTf}
\end{equation}
with $h_F(x)$ is the first derivative of the Fermi-Dirac distribution and $V_0^{\text{f}}(v,\mu_q+xT)$ is the one-loop potential for one fermion field evaluated at $T=0$. This can be written as follows
\begin{align}
    V_0^{(1)\text{f}}(v,\mu_q)&=\frac{m_b^4}{16\pi^2}\Big[ \ln\Big( \frac{4\pi a^2}{(\mu_q+\sqrt{\mu_q^2-m_f^2})^2} \Big)\nonumber \\
    &-\gamma_E+\frac{1}{2}\Big]-\frac{\mu_q\sqrt{\mu_q^2-m_f^2}}{24\pi^2}(2\mu_q^2-5m_f^2).
    \label{V1fLT}
\end{align}

Once again, we notice that the one-loop contribution from fermion fields in the limit $T=0$ that appears in Eq.~(\ref{1lLTf}) is evaluated at $\mu_q \rightarrow \mu_q+xT$. The one-loop contribution for one fermion field in the low temperature approximation then becomes

\begin{align}
    V_{\text{LT}}^{(1)\text{f}}(v,T,\mu_q)=&V_0^{\text{f}}(v,\mu_q+xT)\Big |_{T=0}\nonumber \\
    &+\frac{\pi^2 T^2}{6}\frac{\partial^2}{\partial (xT)^2}V_0^{\text{f}}(v,\mu_q+xT)\Big |_{T=0}\nonumber \\
    &+\frac{\pi^4 T^4}{360}\frac{\partial^4}{\partial (xT)^4}V_0^{\text{f}}(v,\mu_q+xT)\Big |_{T=0}.
    \label{1loopFLT}
\end{align}
For more details see Appendix~D.

Equations~(\ref{treelevel}),~(\ref{1loopBLT}) and~(\ref{1loopFLT}) provide the full expression for the effective potential in the low temperature approximation, which is given by

\begin{align}
    V_{\text{LT}}^{\text{eff}}(v,T,\mu_q,\mu_b)&=-\frac{(a^2+\delta a^2)}{2}v^2
    +\frac{(\lambda+\delta \lambda)}{4}v^4 \nonumber \\
    &-\sum_{i=\sigma,\bar{\pi}}\Big \{ \frac{m_i^4}{64\pi^2}\Big[ \ln \Big( \frac{4\pi^2 a^2}{(\mu_b+\sqrt{\mu_b^2-m_i^2})^2}\Big)\nonumber \\
    &-\gamma_E+\frac{1}{2}  \Big]-\frac{\mu_b\sqrt{\mu_b^2-m_i^2}}{24\pi^2}(2\mu_b^2-5m_i^2)\nonumber \\
    &-\frac{T^2\mu_b}{12}\sqrt{2\mu_b^2-5m_i^2}\nonumber \\
    &-\frac{\pi^2T^4\mu_b}{180}\frac{(2\mu_b^2-3m_i^2)}{(\mu_b^2-m_i^2)^{3/2}} \Big\}\nonumber \\
    &+N_c\sum_{f=u,d}\Big\{ \frac{m_f^4}{16\pi^2}\Big[ \ln \Big( \frac{4\pi^2 a^2}{(\mu_q+\sqrt{\mu_q^2-m_f^2})^2}\Big)\nonumber \\
    &-\gamma_E+\frac{1}{2}  \Big]-\frac{\mu_q\sqrt{\mu_q^2-m_f^2}}{24\pi^2}(2\mu_q^2-5m_f^2)\nonumber \\
    &-\frac{T^2\mu_q}{6}\sqrt{\mu_q^2-m_f^2}\nonumber \\
    &-\frac{7\pi^2T^4\mu_q}{360}\frac{(2\mu_q^2-3m_f^2)}{(\mu_q^2-m_f^2)^{3/2}} \Big\} 
    \label{VeffLT}
\end{align}

\begin{figure}[t]
\begin{center}
\includegraphics[scale=0.48]{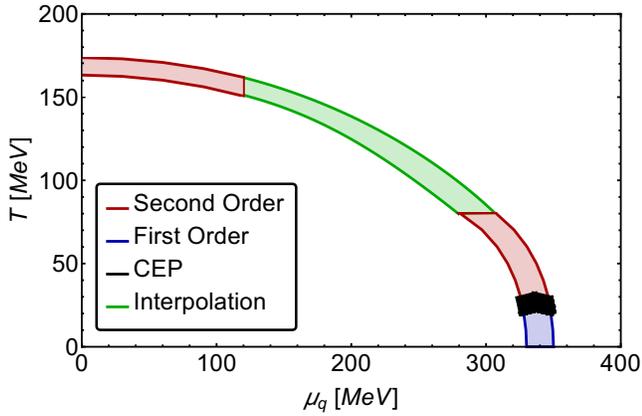}
\end{center}
\caption{QCD phase diagram, using the physical vacuum pion mass, obtained from the solutions to the equations that determine the coupling constants. These are presented in the range $0.77<\lambda <0.86$ and  $1.53<g<1.63$, with $\mu_q=\mu_b$ and the band's upper line computed with $T^c_0(\mu_q=0)=175$ MeV and $\mu^c_q(T=0)=350$ MeV and the lower line with $T^c_0(\mu_q=0)=165$ MeV and $\mu^c_q(T=0)=330$ MeV. The second order transitions are indicated by the shaded red areas and the first order transitions by the blue shaded areas. These areas represent the results directly obtained from our analysis. The intermediate green shaded area is a Pad\'e approximation that interpolates between the high and low temperature regimes.}
\label{fig4}
\end{figure}

We are now in position to explore the QCD phase transition in the regions of the QCD phase diagram where the temperature is larger than the quark chemical potential and where the temperature is smaller than the quark chemical potential. However, before exploring the phase diagram, we need to determine the value of all the parameters involved in the linear sigma model, appropriate for the conditions of the analysis. In the following section we proceed in this direction to determine the values of those parameters and in particular of the couplings $\lambda$ and $g$.

\section{\label{sec:level7}Coupling Constants}

Regardless of the approximation to the effective potential that is being considered, Eq.~(\ref{finalHTpotential}) or Eq.~(\ref{VeffLT}), we observe that we have five free parameters which should be fixed. These are the two coupling constants $\lambda$ and $g$, the square mass parameter $a^2$ and the counter-terms $\delta a^2$ and $\delta\lambda$. In order to determine $a^2$, we use that the vacuum boson masses, Eq.~(\ref{masses}), satisfy

\begin{equation}
    a=\sqrt{\frac{m_\sigma^2-3m_\pi^2}{2}}.
    \label{afixed}
\end{equation}

\begin{figure}[t]
\begin{center}
\includegraphics[scale=0.48]{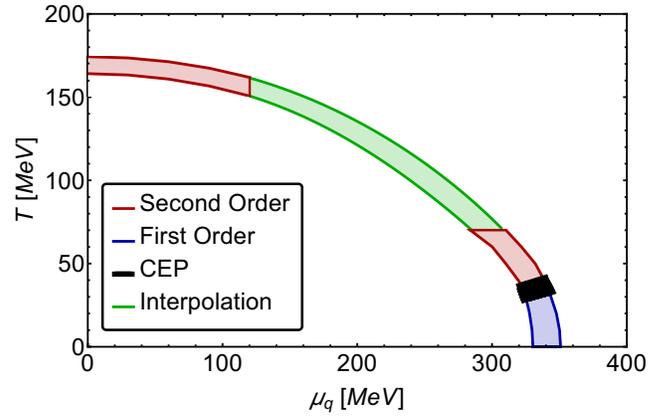}
\end{center}
\caption{QCD phase diagram, using the physical vacuum pion mass, obtained from the solutions to the equations that determine the coupling constants. These are presented in the range $0.45<\lambda <0.49$ and $1.59<g<1.68$, with $\mu_q=2\mu_b$ and he band's upper line computed with $T^c_0(\mu_q=0)=175$ MeV and $\mu^c_q(T=0)=350$ MeV and the lower line with $T^c_0(\mu_q=0)=165$ MeV and $\mu^c_q(T=0)=330$ MeV. The second order transitions are indicated by the shaded red areas and the first order transitions by the blue shaded areas. These areas represent the results directly obtained from our analysis. The intermediate green shaded area is a Pad\'e approximation that interpolates between the high and low temperature regimes.}
\label{fig5}
\end{figure}

\begin{figure}[t]
\begin{center}
\includegraphics[scale=0.48]{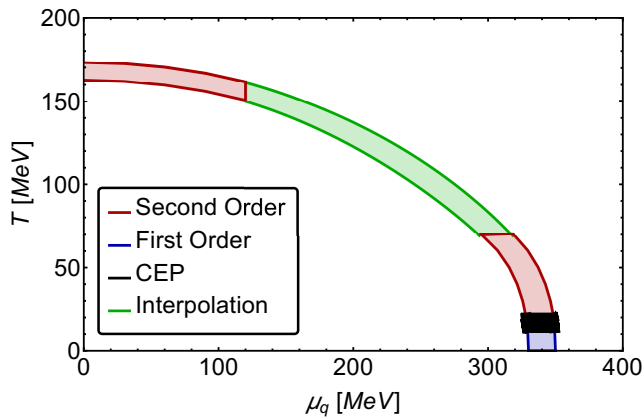}
\end{center}
\caption{QCD phase diagram, using the physical vacuum pion mass, obtained from the solutions to the equations that determine the coupling constants. These are presented in the range $0.99<\lambda <1.10$ and $1.50<g<1.59$, with $\mu_q=0.5\mu_b$ and the band's upper line computed with $T^c_0(\mu_q=0)=175$ MeV and $\mu^c_q(T=0)=350$ MeV and the lower line with $T^c_0(\mu_q=0)=165$ MeV and $\mu^c_q(T=0)=330$ MeV. The second order transitions are indicated by the shaded red areas and the first order transitions by the blue shaded areas. These areas represent the results directly obtained from our analysis. The intermediate green shaded area is a Pad\'e approximation that interpolates between the high and low temperature regimes.}
\label{fig6}
\end{figure}

\begin{figure}[t]
\begin{center}
\includegraphics[scale=0.48]{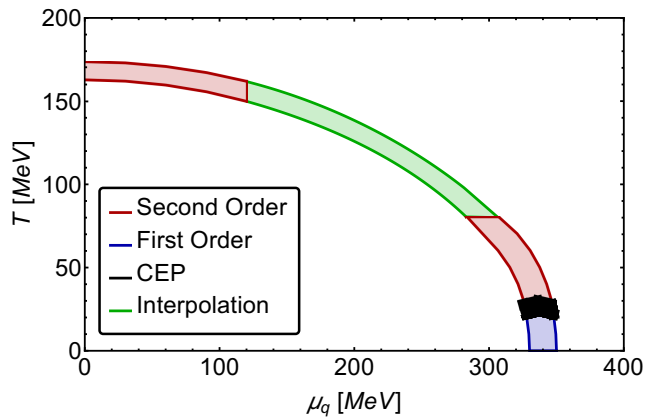}
\end{center}
\caption{QCD phase diagram, in the chiral limit ($m_\pi=0$), obtained from the solutions to the equations that determine the coupling constants. These are presented in the range $1.02<\lambda <1.13$ and $1.78<g<1.89$, with $\mu_q=\mu_b$ and the band's upper line computed with $T^c_0(\mu_q=0)=175$ MeV and $\mu^c_q(T=0)=350$ MeV and the lower line with $T^c_0(\mu_q=0)=165$ MeV and $\mu^c_q(T=0)=330$ MeV. The second order transitions are indicated by the shaded red areas and the first order transitions by the blue shaded areas. These areas represent the results directly obtained from our analysis. The intermediate green shaded area is a Pad\'e approximation that interpolates between the high and low temperature regimes.}
\label{fig7}
\end{figure}

\begin{figure}[t]
\begin{center}
\includegraphics[scale=0.48]{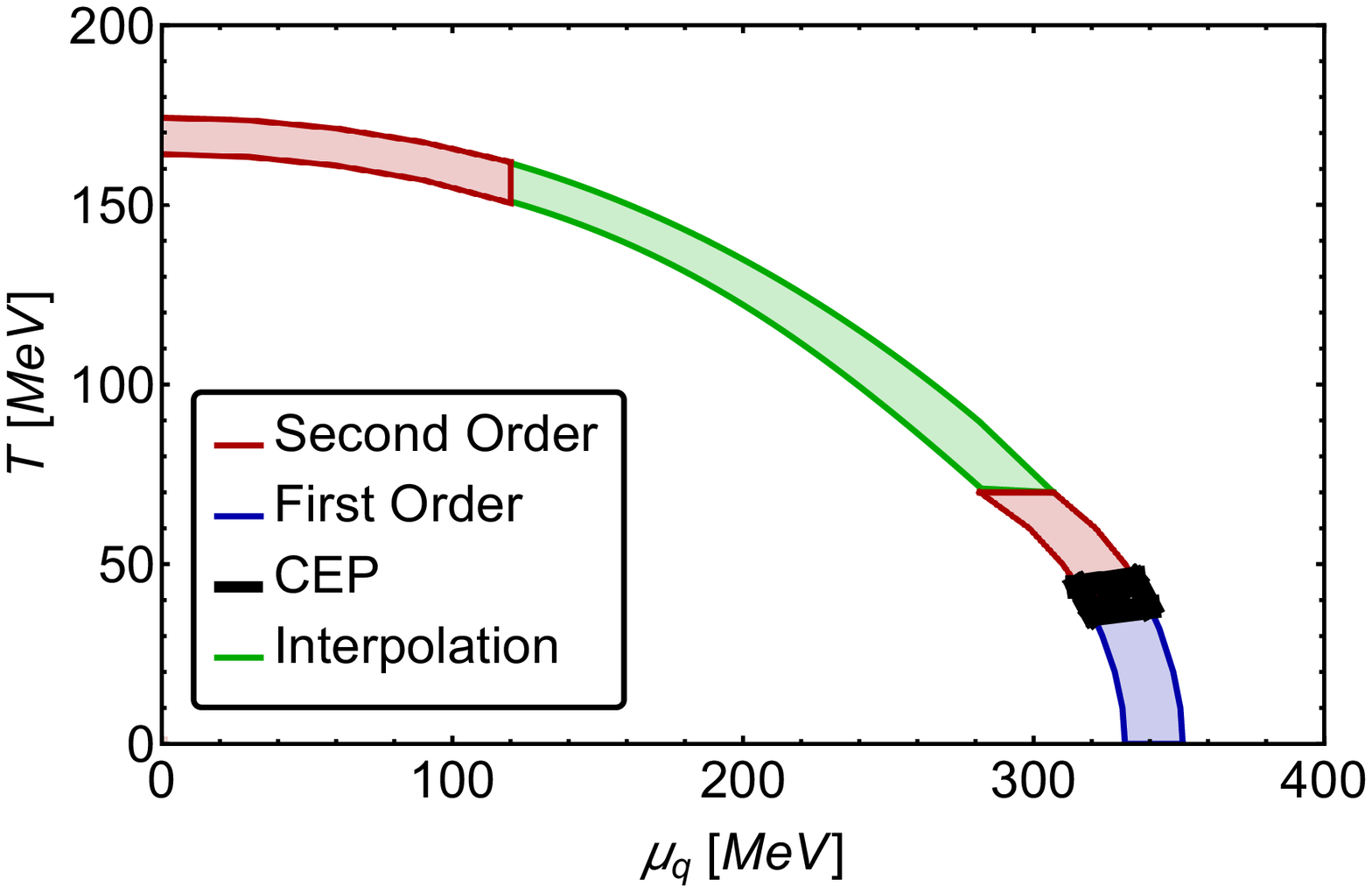}
\end{center}
\caption{QCD phase diagram, in the chiral limit ($m_\pi=0$), obtained from the solutions to the equations that determine the coupling constants. These are presented in the range $0.58<\lambda <0.64$ and $1.84<g<1.96$, with $\mu_q=2\mu_b$ and he band's upper line computed with $T^c_0(\mu_q=0)=175$ MeV and $\mu^c_q(T=0)=350$ MeV and the lower line with $T^c_0(\mu_q=0)=165$ MeV and $\mu^c_q(T=0)=330$ MeV. The second order transitions are indicated by the shaded red areas and the first order transitions by the blue shaded areas. These areas represent the results directly obtained from our analysis. The intermediate green shaded area is a Pad\'e approximation that interpolates between the high and low temperature regimes.}
\label{fig8}
\end{figure}

We can fix $a$ using the physical vacuum sigma and pion masses. This analysis is shown in Figs.~\ref{fig4}-\ref{fig6}. Alternatively, we can work in the strict chiral limit, taking $m_\pi=0$. This analysis is shown in Figs.~\ref{fig7}-\ref{fig9}. We notice that the two kinds of phase diagrams obtained are very similar, in particular the CEP's location changes very little. 

We now need to use two conditions to fix the values of the coupling constants, the main idea is to use physical inputs such that the relations which satisfy $\lambda$ and $g$ are consistent with the realistic behavior of QCD matter around the phase transition in the high and low temperature domains.

From LQCD computations~\cite{latticeTc}, we know that at $\mu_q\equiv\mu_B/3=0$, the QCD phase transition is a crossover, hereby described as a second order transition, and happens for 2+1 light flavors at $T^c_0\simeq 155$ MeV and for only 2 light flavors at $T^c_0\simeq 170$ MeV. In a second order phase transition, the vacuum expectation value ($vev$) continuously transits from the broken phase to the restored phase and thus there is only one minimum. On the other hand, from the analysis using effective models~\cite{models} it is found that at very low values of $T$ and high values of $\mu_q$ the transition is first order. From the analysis based on Hagedorn's limiting temperature~\cite{Hagedorn} at finite $\mu_B$, we know that the critical value for the transition curve to intersect the horizontal axis in the QCD diagram is $\mu_B\simeq m_B$, where $m_B\simeq$ 1 GeV is the typical value of the baryon mass. The $vev$ transits from the broken phase to the restored phase in a discontinuous way. This means that at the phase transition, the effective potential develops two degenerate minima. In one or the other case, the thermal pion mass evaluated at the minima of the potential always vanishes, since this field is a Goldstone mode.

In order to fix the coupling constants we use as inputs the values of temperature and quark chemical potential in two extreme points along the transition curve, namely, when the restoration of chiral symmetry is at $\mu_q=0$ and when it is at $T=0$. Hereafter we refer to these extreme points of the diagram as points ($A$) and ($B$), respectively. 

At point ($A$), the phase transition is second order, hence the square of the pion thermal mass, evaluated at $v=0$ and $T=T_c^0$, is given by

\begin{equation}
 m_\pi^2(0,T^c_0,\mu_q=0)=-a^2+\Pi(T^c_0,\mu_q=0)=0.
 \label{sigmamassT}
\end{equation}
In other words, Eq.~(\ref{sigmamassT}) tells us that the curvature at $v=0$ and $T=T^c_0$ is zero. Therefore the shape of the potential near $v=0$ is flat both in the $\sigma$ and the pion directions. This is depicted in Fig.~\ref{fig2}.

At point ($B$), the phase transition is first order, therefore we expect that at $\mu_q \simeq m_B / 3$ the effective potential develops two degenerate minima. This is depicted in Fig.~\ref{fig3}. Notice that the fermion contribution to the effective potential is responsible for the order of the phase transition. At low densities, this contribution is not strong enough to produce a hump in the effective potential whereas at high densities this contribution produces the barrier between minima at the critical temperature.

Since the analysis we carry out describes the transit from the broken to the restored phase, the minimum we are following is the one with a $vev$ different from zero, which we call $v_1$. This last condition can be written as

\begin{equation}
 m_\pi^2(v_1,0,\mu_q^c)=\lambda v_1-a^2+\Pi(0,\mu_q^c)=0,
 \label{sigmamassMU}
\end{equation}

In Eq.~(\ref{sigmamassMU}), we notice that a new unknown appears: $v_1$, that is, the value of the non-vanishing minimum. The set of conditions necessary to determine all the unknowns is

\begin{align}
 \frac{\partial V^{\text{eff}}}{\partial v}(v=0,T=0,\mu_q=\mu_q^c)&=0, \nonumber \\
 \frac{\partial V^{\text{eff}}}{\partial v}(v=v_1,T=0,\mu_q=\mu_q^c)&=0, \nonumber \\
 V^{\text{eff}}(v=0,T=0,\mu_q=\mu_q^c)&=\nonumber\\
 V^{\text{eff}}(v=v_1,T=0,\mu_q=\mu_q^c)&.
 \label{doubleminima}
\end{align}

The three expressions in Eq.~(\ref{doubleminima}) indicate that the effective potential has two degenerated minima at the phase transition and thus that the transition is first order when $T=0$ and the quark chemical potential is finite and equal to its critical value. 

\section{\label{sec:level8}Results}

The above set of conditions, Eqs.~(\ref{sigmamassT}),~(\ref{sigmamassMU}) and~(\ref{doubleminima}), represent the five algebraic equations that determine the values of $\lambda$ and $g$. These equations provide four pairs of solutions, out of which we pick the pair that corresponds to real positive solutions for $\lambda$ and $g$. We are therefore in the position to explore the QCD phase diagram.

Figures~\ref{fig4} - \ref{fig6} show the phase diagram obtained for the case when the mass parameter $a$ is computed using the physical pion mass in vacuum. These are computed using $\mu_q = \mu_b,\ 2\mu_b,\ 0.5 \mu_b$, respectively. In each figure, the band's upper line is computed with $T^c_0(\mu_q =0)=175$ MeV and $\mu^c_q(T=0)=350$ MeV and the lower line with $T^c_0(\mu_q=0)=165$ MeV and $\mu^c_q(T=0)=330$ MeV. These ranges produce corresponding ranges to the solutions given by ($0.77< \lambda <0.86$,  $1.53<  g<1.63$), ($0.45< \lambda <0.49$, $1.59< g<1.68$), and ($0.99< \lambda <1.10$, $1.50< g<1.59$), respectively. 

\begin{figure}[t]
\begin{center}
\includegraphics[scale=0.48]{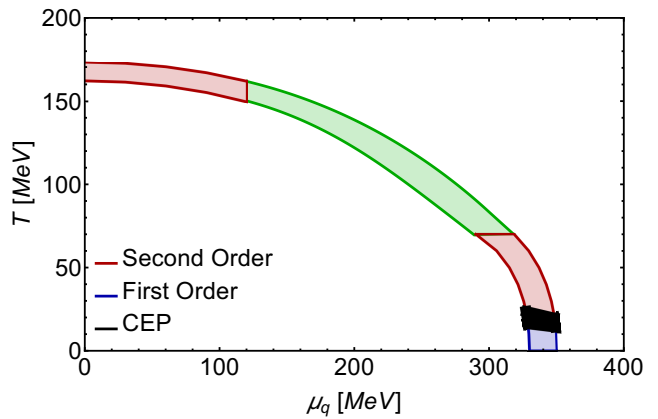}
\end{center}
\caption{QCD phase diagram, in the chiral limit ($m_\pi=0$), obtained from the solutions to the equations that determine the coupling constants. These are presented in the range $1.15<\lambda <1.30$ and $1.74<g<1.85$, with $\mu_q=0.5\mu_b$ and the band's upper line computed with $T^c_0(\mu_q=0)=175$ MeV and $\mu^c_q(T=0)=350$ MeV and the lower line with $T^c_0(\mu_q=0)=165$ MeV and $\mu^c_q(T=0)=330$ MeV. The second order transitions are indicated by the shaded red areas and the first order transitions by the blue shaded areas. These areas represent the results directly obtained from our analysis. The intermediate green shaded area is a Pad\'e approximation that interpolates between the high and low temperature regimes.}
\label{fig9}
\end{figure}

Figures~\ref{fig7} -- \ref{fig9} show the phase diagram obtained for the case when the mass parameter $a$ is computed setting $m_\pi=0$, that is in the chiral limit. These are computed using $\mu_q=\mu_b,\ 2\mu_b,\ 0.5\mu_b$, respectively. In each figure, the band's upper line is computed with $T^c_0(\mu_q=0)=175$ MeV and $\mu^c_q(T=0)=350$ MeV and the lower line with $T^c_0(\mu_q=0)=165$ MeV and $\mu^c_q(T=0)=330$ MeV. These ranges produce corresponding ranges to the solutions given by ($1.02<\lambda <1.13$,  $1.78<g<1.89$), ($0.58<\lambda <0.64$, $1.84<g<1.96$), and ($1.15<\lambda <1.30$, $1.74<g<1.85$), respectively. Notice that the CEP location does not change significantly regardless of weather we set the pion mass either to its physical value or to zero. 

We find that at high (low) temperature and low (high) quark chemical potential the phase transitions are second (first) order. The second order transitions are indicated by the shaded red areas and the first order transitions by the blue shaded areas. These areas represent the results directly obtained from our analysis. The intermediate green shaded area is a Pad\'e approximation that interpolates between the high and low temperature regimes. In all cases, we locate the CEP's region at low temperatures and high quark chemical potential.

\section{\label{sec:level9}Summary and conclusions}

In this work we have used the linear sigma model with quarks to explore the QCD phase diagram from the point of view of chiral symmetry restoration. We have computed the finite temperature effective potential up to the contribution of the ring diagrams to account for the plasma screening effects and have introduced a quark and a boson chemical potentials. The latter is related to the high density of the system that is in turn linked to the high baryon abundance at large values of the quark chemical potential. 

Our approach was to determine the model's couplings using physical inputs such as the vacuum pion and sigma masses, the LQCD value for the critical temperature at $\mu_q=0$ and the conjectured end point value of $\mu_B$ of the transition line at $T=0$. We have also performed the analysis using $m_\pi=0$ instead of its physical vacuum value.  The set of conditions that determine the couplings enforce the requirement that at high temperature the transition is second order whereas at low temperature is first order. Of particular importance is the observation that at the minima of the effective potential, the pion thermal mass vanishes, since this particle is a Goldstone boson.

We find that when varying the values of $T^c_0(\mu_q=0)$ and $\mu^c_q(T=0)$ by 10\% from the chosen central values, the procedure allows to locate the phase transition lines in narrow band. The CEP is however stable when varying the relation between $\mu_q$ and $\mu_b$ and even when the pion mass is set to either its physical value or to zero. The CEP location corresponds to low temperatures and high values of the quark chemical potential. Table~I summarizes the CEP location found in some recent works together with our findings.

\begin{center}
\begin{table}
 \begin{tabular}{||c | c | c||} 
 \hline
       Reference & $T_{CEP}$ & $\mu_{CEP}$\\
 \hline\hline
        C. Shi, \textit{et al.}~\cite{Shi} & 0.85 $T_c$ & 1.11 $T_c$\\
        \hline
        G. A. Contrera, \textit{et al.}~\cite{Contrera}  & 69.9 MeV & 319.1 MeV\\
        \hline
        T. Yokota, \textit{et al.}~\cite{Yokota} & 5.1 MeV & 286.7 MeV \\
        \hline
        S. Sharma~\cite{Sharma} & 145-155 MeV & $>$2 $T_{CEP}$\\
        \hline
        J. Knaute, \textit{et al.}~\cite{Knaute} & 112 MeV & 204 MeV \\
        \hline
        N. G. Antoniou, \textit{et al.}~\cite{Antoniou} & 119-162 MeV & 84-86 MeV \\
        \hline
        Z. F. Cui, \textit{et al.}~\cite{Cui} & 38 MeV & 245 MeV \\
      \hline
       P. Kov\'acs and G. Wolf~\cite{Kovacs} &  & $>$133.3 MeV \\
       \hline
       R. Rougemont, \textit{et al.}~\cite{Rougemont}  & $<$130 MeV& $>$133.3 MeV  \\
       \hline
       This work & 18-45 MeV & 315-349 MeV \\
       \hline
\end{tabular}
\caption{Summary of some recent results for the CEP location, including our results.}
\end{table}
\end{center} 

In order to provide a more robust CEP's location, we need to extend the analytical expansion of the effective potential to a larger temperature range. Perhaps even more important will be to include the temperature and density modifications to the couplings which has been shown useful to describe the inverse magnetic catalysis phenomenon~\cite{inverse}. This is work for the future and will be reported elsewhere.

\section*{Acknowledgments}

Support for this work has been received in part by UNAM-DGAPA-PAPIIT grant number IN101515 and by Consejo Nacional de Ciencia y Tecnolog\'ia grant number 256494.

\section*{Appendices}

\subsection{One-loop effective potential for boson fields. High temperature approximation}
\renewcommand{\theequation}{A\arabic{equation}}
\setcounter{equation}{0}

To compute the vacuum and matter contributions to the boson one-loop effective potential, we start from Eq.~(\ref{V1bwosum}). The vacuum contribution is computed using dimensional regularization. Using the well known expression

\begin{equation}
	\int \frac{d^Dk}{(2\pi)^D} \frac{1}{(k^2-m_b^2)^n}=\text{i}(-1)^n\frac{(m^2)^{2-\epsilon-n}}{(4\pi)^{2-\epsilon}}\frac{\Gamma(n-2+\epsilon)}{\Gamma(n)},
\end{equation}
with $D=d-2\epsilon $, this contribution can be written as

\begin{equation}
	V^{(1)\textrm{b}}_{\textrm{vac}}=\frac{\tilde{\mu}^{3-d}}{2}\int \frac{d^dk}{(2\pi)^d} \sqrt{k^2+m_b^2}.
    \label{vacuumbRD}
\end{equation}

In Eq.~(\ref{vacuumbRD}), we have explicitly $d=3$ and $n=-1/2$. Hence we have

\begin{equation}
	V^{(1)\textrm{b}}_{\textrm{vac}}=-\frac{m_b^4}{32\pi^2}\Gamma(\epsilon-2)\Big( \frac{4\pi \tilde{\mu}^2}{m_b^2} \Big)^\epsilon,
\end{equation}
taking the limit $\epsilon \rightarrow 0$, we finally obtain
\begin{align}
	V^{(1)\textrm{b}}_{\textrm{vac}}=-\frac{m_b^4}{64\pi^2}\Big[ \ln \Big( \frac{4\pi \tilde{\mu}^2}{m_b^2}\Big)
    -\gamma_E+\frac{3}{2}+\frac{1}{\epsilon}\Big].
    \label{finalvaccumB}
\end{align}

We use the Minimal Subtraction scheme (MS). After fixing the renormalization scale to $\tilde{\mu}=a e^{-1/2}$, the final expression for the vacuum contribution is given by

\begin{equation}
V^{(1)\textrm{b}}_{\textrm{vac}}=-\frac{m_b^4}{64\pi^2}\Big[ \ln \Big( \frac{4\pi a^2}{m_b^2}\Big)-\gamma_E+1\Big].
\end{equation}

On the other hand, the matter contribution from one boson field is 

\begin{equation}
	V^{(1)\textrm{b}}_{\textrm{matt}}=\frac{T}{2\pi^2}\int dk \ k^2 \ln \Big(1-e^{-\sqrt{k^2+m_b^2}/T} \Big).
    \label{bosmatter}
\end{equation}
Taking $m_b/T \ll 1$, we can make an expansion of Eq.~(\ref{bosmatter}) in terms of powers of $m_b/T$. The first three terms of the series are given by

\begin{align}
	V^{(1)\textrm{b}}_{\textrm{matt}}=
    -\frac{m_b^4}{64\pi^2}\ln \Big( \frac{m_b^2}{(4\pi T)^2}\Big)-\frac{\pi^2 T^4}{90}+\frac{m_b^2 T^2}{24}-\frac{m_b^3 T}{12\pi}.\nonumber\\
\end{align}
For more details, see Appendix C in Ref.~\cite{DJ}.

\subsection{One-loop effective potential for fermion fields. High temperature approximation}
\renewcommand{\theequation}{B\arabic{equation}}
\setcounter{equation}{0}

The one-loop fermion contribution to the effective potential also contains two terms: the vacuum and matter contributions. The former can be computed following step by step what is done for the boson case. We only notice that the fermion case differs from the boson case by an overall factor $-4$. Therefore if we multiply Eq.~(\ref{finalvaccumB}) by $-4$, we get the vacuum one-loop contribution from one fermion field

\begin{equation}
V^{(1)\textrm{f}}_{\textrm{vac}}=\frac{m_f^4}{16\pi^2}\Big[ \ln \Big( \frac{4\pi a^2}{m_f^2}\Big)-\gamma_E+1\Big].
\end{equation}

The matter contribution from one fermion field is quite similar to the boson case. The two significant differences are (1) fermions obey the Fermi-Dirac distribution and (2) the matter term has particle and anti-particle contributions. This information is encoded in the quark chemical potential. To compute the matter contribution from one fermion, we star from

\begin{align}
    V^{(1)\textrm{f}}_{\textrm{matt}}=
	\frac{T}{\pi^2}\int dk \ k^2&\Big \{
    \ln \Big(1-e^{-(\sqrt{k^2+m_b^2}-\mu_q)/T} \Big) \nonumber \\
    &+\ln \Big(1-e^{-(\sqrt{k^2+m_b^2}+\mu_q)/T} \Big)\Big \}.
\end{align}
In the high temperature approximation, $m_f/T\ll 1$ and $\mu_q/T<1$. This approximation allows us to explore the  phase diagram's region where the temperature is larger than baryon chemical potential. Proceeding in a fashion entirely analogous to the boson case, one obtains
\begin{eqnarray}
	V^{(1)\textrm{f}}_{\textrm{matt}}&=&
    \frac{m_f^4}{16\pi^2}\Big[\ln \Big( \frac{m_f^2}{(4\pi T)^2}\Big)\nonumber\\
    &-&\psi^0\Big( \frac{1}{2}+\frac{\text{i}\mu}{2\pi T} \Big)\nonumber \\
    &-&\psi^0\Big( \frac{1}{2}-\frac{\text{i}\mu}{2\pi T} \Big)\Big]-8m_f^2T^2\Big[ \text{Li}_2(-e^{\mu_q/T})\nonumber \\
    &+&\text{Li}_2(-e^{-\mu_q/T}) \Big]+32T^4\Big[ \text{Li}_4(-e^{\mu_q/T})\nonumber \\
    &+&\text{Li}_4(-e^{-\mu_q/T}) \Big].
    \label{matterF}
\end{eqnarray}
For more details, see Appendix C in Ref.~\cite{DJ}.

\subsection{One-loop effective potential for boson fields. Low temperature approximation}
\renewcommand{\theequation}{C\arabic{equation}}
\setcounter{equation}{0}

In the low temperature approximation, we work in the regime where the quark chemical potential is the most important energy scale and temperature is the smallest one. In order to obtain the one-loop contribution to the effective potential from one boson field, we expand Eq.~(\ref{V1bwosum}) in a Taylor series around $T=0$. For this purpose, we start from Eq.~(\ref{1lLT}), namely

\begin{align}
	\int_{\frac{\mu_b-m_b}{T}}^\infty &V_0^{(1)\text{b}}(v,\mu_b+xT)h_B(x)dx=\nonumber \\
    & V_0^{(1)\text{b}}(v,\mu_b+xT)\Big |_{T=0} \int_{\frac{\mu_b-m_b}{T}}^\infty h_B(x)dx\nonumber \\
    &+\frac{\partial^2 ( V_0^{(1)\text{b}}(v,\mu_b+xT))}{\partial (xT)^2}\Big |_{T=0} \int_{\frac{\mu_b-m_b}{T}}^\infty x^2h_B(x)dx \nonumber \\
    &+\frac{\partial^4 ( V_0^{(1)\text{b}}(v,\mu_b+xT))}{\partial (xT)^4}\Big |_{T=0} \int_{\frac{\mu_b-m_b}{T}}^\infty x^4h_B(x)dx \nonumber \\
    &+ \cdots \ ,
    \label{TaylorB}
\end{align}
with
\begin{align}
	\int_{\frac{\mu_b-m_b}{T}}^\infty h_B(x)dx&=1 \nonumber \\
    \int_{\frac{\mu_b-m_b}{T}}^\infty x^2h_B(x)dx&=\frac{\pi^2 T^2}{12} \nonumber \\
    \int_{\frac{\mu_b-m_b}{T}}^\infty x^4h_B(x)dx&=\frac{7\pi^4T^4}{1260}.
    \label{coefficientsB}
\end{align}
Therefore, in the low temperature approximation, the one-loop contribution from one boson field is given by
\begin{eqnarray}
    V^{(1)\text{b}}_{\text{LT}}(v,T,\mu_b)&=&V_0^{(1)\text{b}}(v,\mu_b)+\frac{\pi^2 T^2}{12}\frac{\partial^2}{\partial T^2}V_0^{(1)\text{b}}(v,\mu_b) \nonumber \\
    &+&\frac{7\pi^4 T^4}{1260}\frac{\partial^4}{\partial T^4}V_0^{(1)\text{b}}(v,\mu_b).
\end{eqnarray}
\vspace{0.5cm}

\subsection{One-loop effective potential for fermion fields. Low temperature approximation}
\renewcommand{\theequation}{D\arabic{equation}}
\setcounter{equation}{0}

Following a procedure in a fashion entirely similar to the boson case, we make a Taylor expansion around $T=0$. The one-loop contribution from one fermion field in the low temperature approximation becomes

\begin{align}
	\int_{\frac{\mu_q-m_f}{T}}^\infty &V_0^{(1)\text{f}}(v,\mu_q+xT)h_F(x)dx=\nonumber \\
    & V_0^{(1)\text{f}}(v,\mu_q+xT)\Big |_{T=0} \int_{\frac{\mu_q-m_f}{T}}^\infty h_F(x)dx\nonumber \\
    &+\frac{\partial^2 ( V_0^{(1)\text{f}}(v,\mu_q+xT))}{\partial (xT)^2}\Big |_{T=0} \int_{\frac{\mu_q-m_f}{T}}^\infty x^2h_F(x)dx \nonumber \\
    &+\frac{\partial^4 ( V_0^{(1)\text{f}}(v,\mu_q+xT))}{\partial (xT)^4}\Big |_{T=0} \int_{\frac{\mu_q-m_f}{T}}^\infty x^4h_F(x)dx \nonumber \\
    &+ \cdots \ ,
    \label{TaylorF}
\end{align}
where we have substituted $m_b\rightarrow m_f$, $\mu_b \rightarrow \mu_q$ and $h_B(x) \rightarrow h_F(x)$. We now obtain that

\begin{align}
	\int_{\frac{\mu_q-m_f}{T}}^\infty h_F(x)dx&=1 \nonumber \\
    \int_{\frac{\mu_q-m_f}{T}}^\infty x^2h_F(x)dx&=\frac{\pi^2 T^2}{6} \nonumber \\
    \int_{\frac{\mu_q-m_f}{T}}^\infty x^4h_F(x)dx&=\frac{\pi^4T^4}{360}.
    \label{coefficientsF}
\end{align}

Therefore, substituting Eq.~(\ref{coefficientsF}) into Eq.~(\ref{TaylorF}), we finally get Eq.~(\ref{1loopFLT}).

\end{document}